# COVID-19Base: A knowledgebase to explore biomedical entities related to COVID-19

Junaed Younus Khan[1], Md. Tawkat Islam Khondaker[1], Iram Tazim Hoque[1], Hamada Al-Absi[2], Mohammad Saifur Rahman[1], Tanvir Alam[2,*], M. Sohel Rahman[1]

[1]Department of Computer Science and Engineering, Bangladesh University of Engineering and Technology, Dhaka, Bangladesh.

[2]Information and Computing Technology Division, College of Science and Engineering, Hamad Bin Khalifa University, Doha, Qatar.

*To whom correspondence should be addressed.

## Abstract

We are presenting COVID-19Base, a knowledgebase highlighting the biomedical entities related to COVID-19 disease based on literature mining. To develop COVID-19Base, we mine the information from publicly available scientific literature and related public resources. We considered seven topic-specific dictionaries, including human genes, human miRNAs, human lncRNAs, diseases, Protein Databank, drugs, and drug side effects, are integrated to mine all scientific evidence related to COVID-19. We have employed an automated literature mining and labeling system through a novel approach to measure the effectiveness of drugs against diseases based on natural language processing, sentiment analysis, and deep learning. To the best of our knowledge, this is the first knowledgebase dedicated to COVID-19, which integrates such large variety of related biomedical entities through literature mining. Proper investigation of the mined biomedical entities along with the identified interactions among those, reported in COVID-19Base, would help the research community to discover possible ways for the therapeutic treatment of COVID-19.

**Availability:** COVID-19Base is available at: http://77.68.43.135:96/.

**Contact:** talam@hbku.edu.qa

## Introduction

Severe acute respiratory syndrome coronavirus 2 (SARS-CoV-2) initially spread widely in China, then in Italy, and, now in other parts of the world, causing coronavirus disease 2019 (COVID-19) [1]. SARS-CoV-2 is the novel coronavirus that causes the alarming pandemic of COVID-19 since February 2020 [2]. While the novel coronaviruses (SARS-CoV-2) have gained enormous attention and grave concern as a consequence of the current worldwide COVID-19 pandemic situation, other known human coronaviruses, beta coronaviruses (SARS-CoV, MERS, OC43, HKU1) and alpha coronaviruses (229E, NL63), historically, caused a significant level of public-health concerns and resulted in severe respiratory syndrome for the patients as well [3, 4]. To combat the global lethality of COVID-19, it demands an urgent solution for the detection and therapeutic treatment, which requires a comprehensive experimental elucidation of biomedical entities (e.g., genes, non-coding RNAs (ncRNA), viruses, drugs, etc.) [5] that are involved in this disease. But this is a relatively slow process due to the inherent nature of the experimental validation. As an alternative, *in silico* methods are also proposed by multiple research groups [6, 7]. Still, it is also a daunting task due to a large number of possible combinations of biomedical entities that need to be examined [8]. To mediate these two extreme experimental setups, and at the same time, to enable a comprehensive exploration of potential therapeutic treatments, knowledgebase based solutions are proposed, as part of this study, for the scientific community to focus on a smaller number of biomedical entities that may guide to the discovery of a novel solution of the COVID-19 treatment.

There exist databases that focused on the virus-related diseases for multiple hosts. For example, in ViRBase [9], the authors highlighted the association between non-coding RNAs (ncRNAs) and viruses in 20 hosts. VISDB database, based on literature mining, integrated the virus interaction site in humans for five DNA oncoviruses and four RNA retroviruses [10]. The Virus Pathogen Resources (VIPR), developed a portal collecting a comprehensive set of information related to coronavirus and hepatitis C virus (HCV), other viruses [11, 12] . But none of the mentioned databases are much useful for COVID-19/SARS-CoV-2, as those databases were not specific to novel coronavirus, or they provide very limited information about the associated genes, ncRNA or do not cover other associated factors that are involved in coronavirus-related diseases, drugs and their side effects. Moreover, there exists no such knowledgebase that has integrated all biomedical entities specific to COVID-19/SARS-CoV-2. To fill this gap, we develop the first comprehensive knowledgebase, COVID-19Base, that integrates several biomedical entities that are associated with COVID-19/SARS-CoV-2 from the scientific literature.

In COVID-19Base, we use a dictionary-based approach to find the association of seven different thematic areas related to COVID-19/SARS-CoV-2 and other coronavirus-related diseases in humans. In COVID-19Base, we not only highlight the biomedical terms that are mentioned in the literature, but we also identify "term pairs" based on their co-occurrence, which will allow the scientific community to investigate in depth the association between term pairs like disease-gene, disease-drug, disease-ncRNA, etc. While careful manual curation of the identified associations is the ultimate goal, in COVID-19Base, we propose and implement a novel approach to estimate the effectiveness of drug for diseases based on natural language processing, sentiment analysis, and deep learning. We also apply the concept of cosine similarity [36] to confidently infer the associations between diseases and genes, lncRNAs, miRNAs. The knowledgebase, we developed, will support the researcher around the world to discover the existing knowledge and find a solution for this pandemic.

## Materials and Methods

For this study, we considered COVID-19 Open Research Dataset (CORD-19), generated by Allen Institute of AI (publicly available at https://pages.semanticscholar.org/coronavirus-research) . The dataset contains over 44K scholarly articles, related to COVID-19 and the coronavirus family of viruses. The dataset was collected using query: "COVID-19" OR Coronavirus OR "Corona virus" OR "2019-nCoV" OR "SARS-CoV" OR "MERS-CoV" OR "Severe Acute Respiratory Syndrome" OR "Middle East Respiratory Syndrome" against PubMed, PubMed Central (PMC), bioRxiv and medRxiv pre-prints and it covers major part of all research articles related to COVID-19 and other coronavirus (e.g. MERS, SARS, etc.) till 27-03-2020. We considered the abstract and full body from the research articles for downstream analysis.

**Source of dictionaries:** Genes (HGNC), miRNA (miRBase), lncRNA(LNCipedia), PDB, disease (DO) , drug (DrugBank), side effect (SIDER)

# Extracting Disease-Drug Interactions

We extract disease-drug interactions from the CORD-19 literature and classify them in two categories (labels): Positive and Negative. The positive label means the drug is somewhat effective for curing the disease, and the negative label means the opposite. We also determine a confidence score which indicates our level of confidence for that automatic label. Fig. 1 shows the workflow of extracting disease-drug interactions and predicting the effectiveness of drugs against diseases with confidence scores.

## Disease and Drug Names Extraction

To extract relevant disease-drug pairs from the CORD-19 literature, we employ a dictionary-based approach to detect mentions of diseases and drugs in the literature. We use Disease Ontology [13] and DrugBank [14] to prepare the disease and drug dictionaries. We leverage the Aho-Corasick algorithm [15] to search the drug and disease names considering the large size of drug and disease dictionaries and the corpus itself. The Aho-Corasick algorithm is a string searching algorithm that efficiently locates multiple patterns in a large blob of text. The time complexity of the algorithm is $O(n + m + z)$, where $n$ is the length of the text, $m$ is the total length of all the patterns to be searched, and $z$ is the total number of occurrences of the patterns in the text.

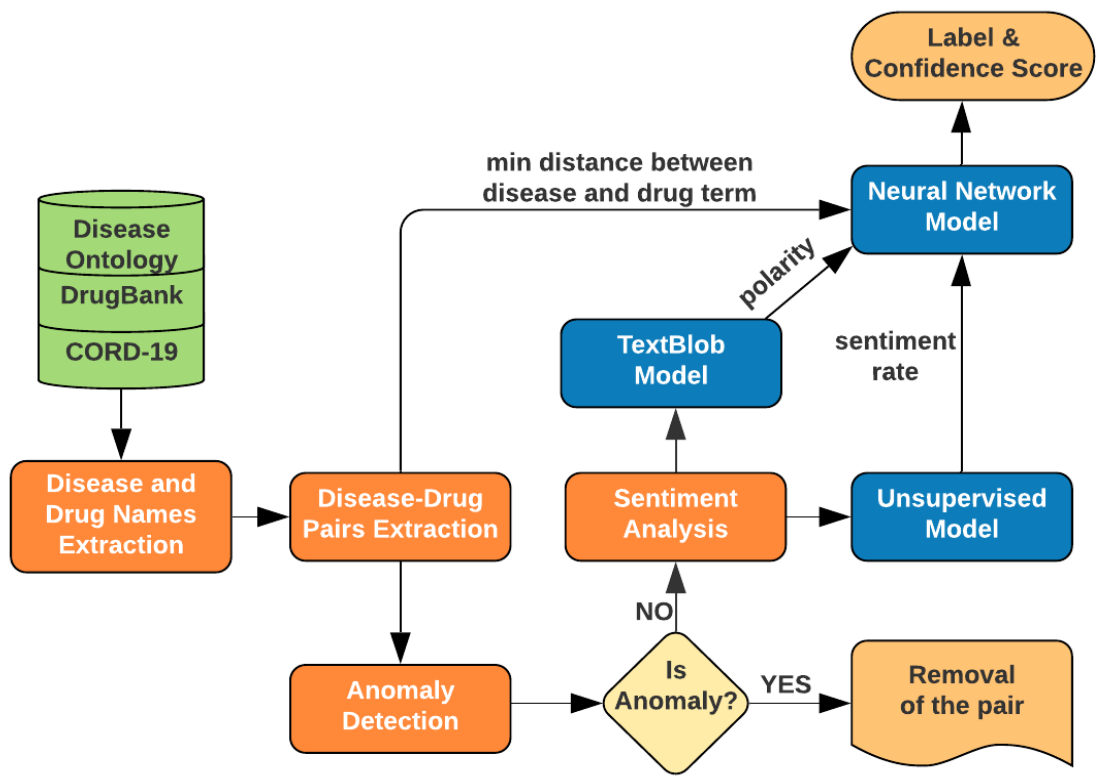

**Figure 1.** Flowchart of extracting disease-drug interactions and predicting the effectiveness of drugs against diseases with confidence scores.

## Disease-Drug Pairs Extraction

After extracting the disease and drug names separately, we want to mine the literature and identify the sentences that contain the disease and drug pairs to semantically evaluate their interactions. For this purpose, we search for every disease-drug pair from our disease and drug list in the CORD-19 literature and collect every sentence where a co-occurrence is found. Then we create a document for every disease-drug pair combining all extracted sentences. Thus we build a disease-drug pair to document mapping. We do not use any pattern-based approach here as done in the paper [16] as this could result in missing a good number of sentences containing disease-drug pairs.

**Anomaly Removal**

As we automatically extract the sentences containing the disease-drug pairs, there is a possibility of aberration in our extracted data. So we decide to check and remove any abnormality from our collected data before going into the next stage of the pipeline. We use unsupervised anomaly detection [17] for doing this job. Unsupervised anomaly detection technique detects anomalies in an unlabeled dataset by looking for instances that seem to fit the least to the remainder of the dataset under the assumption that the majority of the instances in the dataset are 'normal'. We use K-Means clustering algorithm [18] as it has been used for anomaly detection in several studies [19, 20, 21, 22, 23]. We proceed as follows. First, we use Doc2Vec [24] to create a numeric representation of each document associated with each disease-drug pair. Then we fit these representations into our K-means model and observe two clear clusters of easily discriminable sizes, where the smaller one consists of only 161 instances. As we know that anomalies differ from the normal instances significantly and occur very rarely in the data, we can assume that the instances of the smaller cluster are indeed anomalies. We also check a number of instances manually to verify our assumption. We discard these 161 instances from any further consideration.

**Sentiment Analysis**

We apply sentiment analysis to automatically assess the effectiveness of a drug to treat a particular disease in the context of each extracted drug-disease pair. First, we apply the concept of transfer learning. We use TextBlob [25] which is a pre-trained sentiment analysis tool provided as a Python library. However, it shows some inconsistency in some cases as expected from a pre-trained model and we feel the necessity of unsupervised sentiment analysis which is the second model of our pipeline. We get a polarity score from the TextBlob model and a sentiment rate from our unsupervised model for each disease-drug pair which are subsequently fed to our Neural Network model to predict the final label.

**TextBlob model:** TextBlob is a Python library that is widely used in natural language processing tasks such as POS tagging, noun phrase extraction, sentiment analysis, classification, translation, and more. Given the sentences that we mined for each disease-drug pair as input, TextBlob gives a polarity score in the range between -1 and 1. We record the polarity scores for each disease-drug to use it as a feature for our Neural Network model.

**Unsupervised model:** We use the concept of K-means clustering again for unsupervised sentiment analysis. First, we train the Word2Vec [26] model with our mined literature and get a vector representation of every word. Then we run K-means clustering on the estimated word vectors and find two clusters (positive and negative). The positive cluster is decided on the basis of the presence of several positive words (in the context of a disease-rug pair) in it such as 'cure', 'preclude', 'inhibit', 'prescribe', 'reduce', 'modest', etc. On the other hand, the negative cluster contains words like 'risky', 'kill', 'danger', etc. Then we assign each word a sentiment value, either +1 or -1, based on the cluster (positive or negative) they belong to. We weigh this value by dividing it by the distance between the word and the centroid of its cluster to describe the extent of its potential positive or negative-ness. Then we calculate the tf-idf score [27, 28] of each word in the sentence collection to consider the significance of the unique words. Next, we build a tf-idf representation, $T$, for each disease-drug pair by replacing each word of the corresponding sentences with its tf-idf score and a sentiment value representation, $S$, by replacing each word with its sentiment value. Finally, we take their dot product ($T \circ S$) as the final sentiment rate of our unsupervised model.

## Neural Network Model for Automatic Label and Confidence Score

We use a Deep Neural Network (DNN) model to automatically predict the label and confidence score for our disease-drug pairs. Earlier studies [29, 30] also used DNN models on small labeled data and achieved promising results.

**Training data:** We manually labeled 200 disease-drug pairs to train our Neural Network model. Among them, there were 110 positive instances and the rest were negative.

**Input features:** We use the polarity or sentiment score given by the TextBlob and unsupervised models as the input features for our Neural Network model along with the minimum distance between the disease and drug term in the corresponding document.

**Model setup and output:** The DNN structure used in this study is similar to that shown in Fig. 2. It consists of 1 input layer with 3 neurons (each neuron corresponds to one input feature), 2 hidden layers with 8 and 4 neurons respectively, and 1 output layer containing 1 neuron for binary classification (positive or negative). The transfer functions of the first and second hidden layers are Rectified Linear Unit (ReLU) [31] and Hyperbolic Tangent function (tanh) [32] respectively. The transfer function of the output layer is a Sigmoid function [33]. We train the DNN model using Xavier initialization [34] which tries to make the variance of the outputs of a layer to be equal to the variance of its inputs. We use Adam optimizer [35] and the maximum training epoch is set to 500. We split our labeled data into training and test sets on an 80:20 ratio. We train our model on the training data and achieve 75% accuracy on the test set.

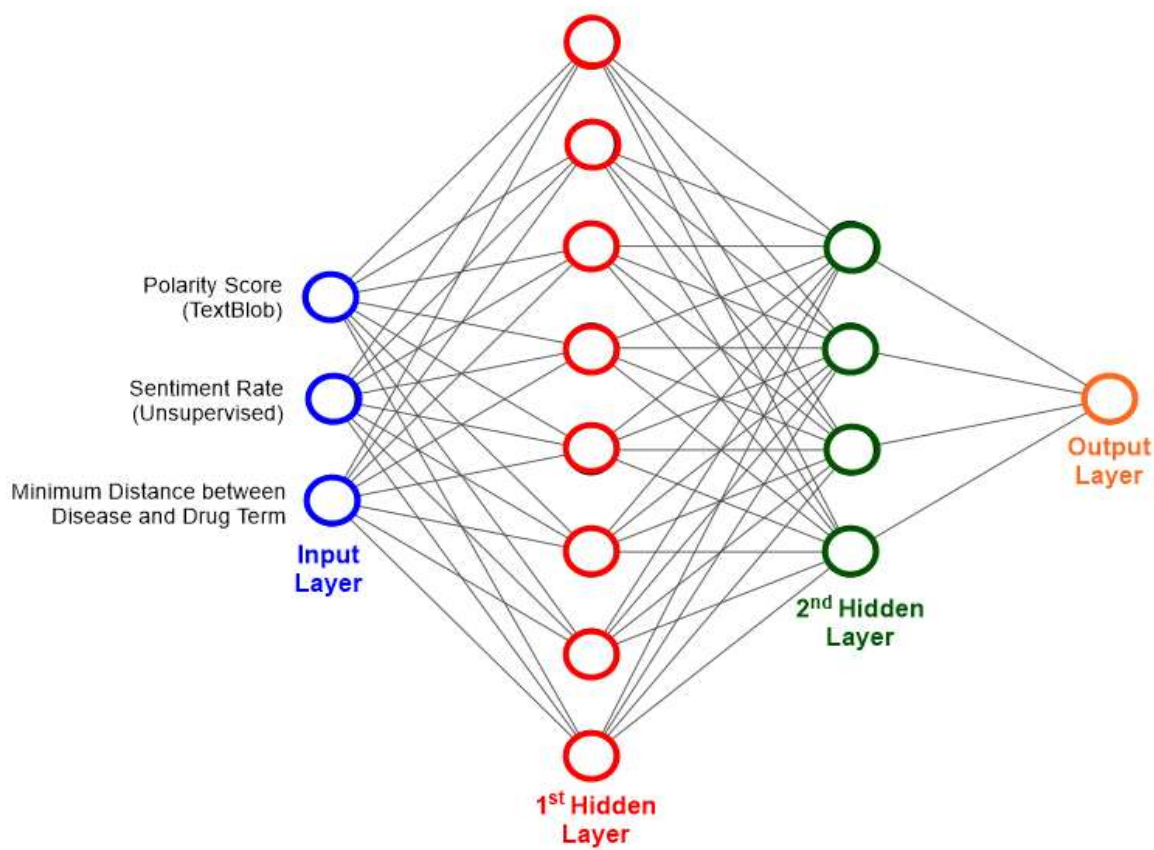

**Figure 2.** Schematic diagram of the deep neural network used to predict the effectiveness of drugs against diseases.

## Extracting Disease-Gene Associations

Fig. 3 shows the workflow of extracting disease-gene associations. We extract gene names along with lncRNAs and miRNAs from the CORD-19 literature in a dictionary-based approach using HGNC [40], LNCipedia [41], and miRBase [42]. Then we extract their associations with diseases in a similar process that we have used to extract the disease-drug pairs and collected all the abstracts where a co-occurrence was found. Next, we apply the concept of cosine similarity [36] to confidently infer the associations. We transformed each disease into vector V1, each gene (and lncRNA, miRNA) into vector V2, and then calculated the cosine similarity of V1 and V2 for each pair. To create the vector representations, we train a Word2Vec model with all the collected abstracts. We use the DisGeNET [37] database as the gold standard to evaluate the performance of cosine similarity in predicting the gene-disease linkage. First, we calculate the maximum, average, and minimum cosine similarity of the pairs that are common both in our findings and in the DisGeNET database. We find that 99.7% of the newly discovered pairs lie within this range (in terms of cosine similarity). We further classify the associations in three classes (high, medium, and low) in terms of confidence as follows: pairs having cosine similarity closest to the maximum (minimum) of the known ones are considered as high (low) confidence associations, and the remaining ones (closest to the average) as medium confidence associations. Moreover, pairs that are also found in the DisGeNET database are labeled as verified associations.

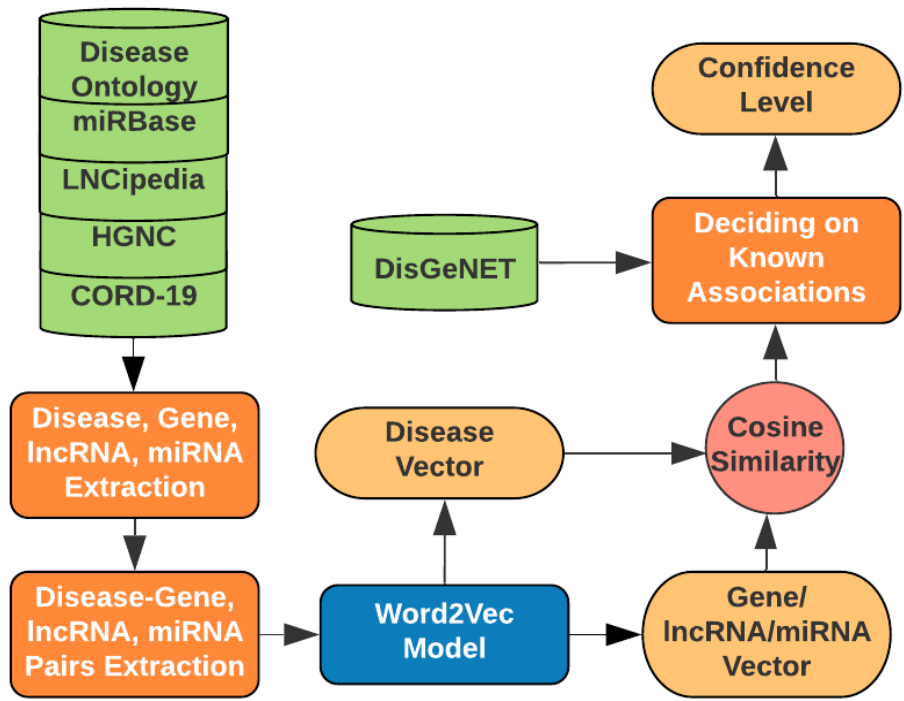

**Figure 3.** Flowchart of extracting disease-gene, disease-lncRNA, disease-miRNA associations and determining their confidence levels.

## Extracting Drug-Protein Associations

We also extract drug-protein associations from the CORD-19 literature applying the same co-occurrence based approach as mentioned above. We use PDB Ids from the Protein Data Bank [38] for extracting protein names. Unlike the disease-gene associations, we do not apply the concept of cosine similarity here as we do not find any suitable dataset that can be used as the gold standard in this case.

## Extracting Side-effects of Drugs

The drugs we are suggesting through this literature mining may come with different side-effects. Therefore, we also explore the possible side-effects of the drugs. We collect the drugs with the corresponding side-effects from SIDER [39] and map the drug names with the drugs mentioned in the CORD-19 literature to extract the possible side-effects.

# Results

## Association between pair of terms in COVID-19Base

We have mined 1811 drugs, 1219 diseases, 1703 genes, 11 lncRNAs, 9 miRNAs, and 70 PDBs from the CORD-19 literature. Table 1 summarizes the interactions or associations between different entities. Among the disease-drug pairs, 11480 are found to be positive and 1273, negative. Among the disease-gene associations, 1977 are verified (V), 40 associations are found with high-confidence (H), 15509 with medium-confidence (M), and 7771 with low-confidence (L). More results are available in Table 1. Majority share of the data is labeled automatically following the methodology reported in the Methods section and complete manual curation is planned in subsequent versions. Notably, a tiny part (1.5%) of the findings have already been manually curated.

**Table 1. Pairs of terms as identified in the analyzed set of documents. Here V, H, M and L means Verified, High-, Medium- and Low-confidence associations respectively. +ve (-ve) indicates an (not) effective association.**

| Interaction/Association | # of extracted pairs of terms |
|---|---|
| Disease-Drug | 12753 (11480+ve, 1273-ve) |
| Disease-Gene | 25297 (1977 V, 40 H, 15509 M, 7771 L ) |
| Disease-lncRNA | 49 ( 2 V, 1 H, 30 M, 16 L ) |
| Disease-miRNA | 57 ( 29 M, 28 L ) |
| Drug-PDB | 288 |

In the rest of this section, we discuss two of the interesting and useful findings using COVID-19Base in the context of drug exploration for COVID-19.

## Remdesivir as the top ranked drug for COVID-19 in COVID-19Base

As a case study, we find Remdesivir as a positive (i.e., effective) drug for COVID-19, automatically labeled as such through our pipeline with a confidence score of 75.86%. Thus COVID-19Base suggests Remdesivir as a promising drug for further investigation for treating COVID-19. Interestingly, it is recently being considered as the most effective drug for treating COVID-19. Notably, Remdesivir is an antiviral drug originally developed for Ebola treatment. It prevents viral replication by blocking an enzyme called polymerase and thus limits the spread of the disease. A recent clinical trial conducted by the National Institute of Allergy and Infectious Diseases (NIAID) shows that Remdesivir helps COVID-19 patients recover faster and improves their survival rates. Adult patients, treated with Remdesivir, were found to recover four days faster, an improvement of 31% compared to other patients and the overall death rate dropped from 11.6% to 8%.[1]

## Hydroxychloroquine as a possible treatment for COVID-19

Anti-malaria drug Hydroxychloroquine, which is one of the most talked-about drugs for treating COVID-19, has also been found in our mining. Our model finds it as a positive (i.e., effective) drug with 57.89% confidence (note the significantly lower confidence score than that of Remdesivir). But it is also revealed from COVID-19Base that this drug has 111 side-effects including Anaemia, Haemorrhage, Liver disorder, Hepatitis fulminant, Cardiomyopathy, Cardiac failure, etc., which makes it a risky option especially for patients with heart and liver complications. Informatively, although the US Food and Drugs Administration (FDA) had previously granted authorization to use this drug for Covid-19, it recently cautions against its use outside of the hospital setting or a clinical trial due to its' side effects and risk factors.[2]

## Availability

COVID-19Base is available at: http://77.68.43.135:96/ . As the number of scientific publications, particularly, on COVID-19 is surging, we will update the knowledgebase on a monthly basis and integrate all the recent updates in the knowledgebase. Additionally, a manually curation drive is underway and to facilitate that a limited sentence-wise feedback mechanism (from the esteemed users) is also incorporated in the software.

## Limitations

While complete manual curation is the ideal goal, due to time and cost constraints we have employed a novel sophisticated automated approach to prepare COVID-19Base and make it available for the scientific community. However, this comes naturally with some errors due to inherent limitations of the methods and approaches adopted. This is why the identified inferences/associations are made available to the users for review to facilitate a feedback mechanism (from the esteemed users) in the COVID-19Base.

---

[1] https://www.niaid.nih.gov/news-events/nih-clinical-trial-shows-remdesivir-accelerates-recovery-advanced-covid-19

[2] https://www.fda.gov/drugs/drug-safety-and-availability/fda-cautions-against-use-hydroxychloroquine-or-chloroquine-covid-19-outside-hospital-setting-or